\documentclass[prb,11pt,preprint]{revtex4-1}

\usepackage{graphicx}			% Include figure files
\usepackage{dcolumn}			% Align table columns on decimalpoint
\usepackage{bm}					  % bold math
\usepackage{hyperref}			% add hypertext capabilities

\usepackage{comment}
\usepackage{amsmath}
\usepackage{cleveref}
\usepackage{amsfonts}
\usepackage{amssymb}
\usepackage{pstcol,times,color,graphicx,graphics,pstricks,pst-node,pst-coil,pst-grad}

\newcommand{\ket}[1]{\ensuremath{\left|{#1}\right\rangle}} % comando para kets
\begin{document}

\preprint{PREPRINT VERSION 1}

\title{
%Squeezing in a harmonic oscillator with two sudden frequency changes: an algebraic solution.
%Two jumps in frequency of a quantum harmonic oscillator: an algebraic solution
A time-dependent harmonic oscillator with two frequency jumps: an exact algebraic solution} 

% % % % % % % % % % % % % % % % % % % % % %

\author{D. M. Tibaduiza}
 \email{Correspondence to: danielmartinezt@gmail.com}
\affiliation{Instituto de F\'{\i}sica, Universidade Federal do Rio de Janeiro \\
 	Avenida Athos da Silveira Ramos, 149, Centro de Tecnologia, Bloco A, Cidade Universit\'aria, Rio de Janeiro-RJ, Brazil\\
 	CEP: 21941-972 Caixa Postal 68528 }
	
\author{L.~Pires}
\affiliation{Instituto de F\'{\i}sica, Universidade Federal do Rio de Janeiro \\
 	Avenida Athos da Silveira Ramos, 149, Centro de Tecnologia, Bloco A, Cidade Universit\'aria, Rio de Janeiro-RJ, Brazil\\
 	CEP: 21941-972 Caixa Postal 68528 }

\author{D.~Szilard}
\affiliation{Instituto de F\'{\i}sica, Universidade Federal do Rio de Janeiro \\
 	Avenida Athos da Silveira Ramos, 149, Centro de Tecnologia, Bloco A, Cidade Universit\'aria, Rio de Janeiro-RJ, Brazil\\
 	CEP: 21941-972 Caixa Postal 68528 }
	
\author{C.~A.~D.~Zarro}
 \affiliation{Instituto de F\'{\i}sica, Universidade Federal do Rio de Janeiro \\
 	Avenida Athos da Silveira Ramos, 149, Centro de Tecnologia, Bloco A, Cidade Universit\'aria, Rio de Janeiro-RJ, Brazil\\
 	CEP: 21941-972 Caixa Postal 68528 }

 \author{C.~Farina }
\affiliation{Instituto de F\'{\i}sica, Universidade Federal do Rio de Janeiro \\
	Avenida Athos da Silveira Ramos, 149, Centro de Tecnologia, Bloco A, Cidade Universit\'aria, Rio de Janeiro-RJ, Brazil\\
	CEP: 21941-972 Caixa Postal 68528 }

\author{A. L. C. Rego}
 \affiliation{Instituto de Aplica\c{c}\~{a}o Fernando Rodrigues da Silveira, Universidade do Estado do Rio de Janeiro \\
 Rua Santa Alexandrina, 288, Rio de Janeiro-RJ, Brazil\\ 
  CEP: 20261-232 }

% % % % % % % % % % % % % % % % % % % % % % 

\date{\today}

\begin{abstract}

We consider a harmonic oscillator (HO) with a time dependent frequency which undergoes two successive abrupt changes. By assumption, the HO starts in its fundamental state with frequency $\omega_0$, then, at $t=0$, its frequency suddenly increases to $\omega_1$ and, after a finite time interval $\tau$, it comes back to its original value $\omega_0$. Contrary to what one could naively think, this problem is a quite non-trivial one. Using algebraic methods we obtain its exact analytical solution and show that at any time $t>0$  the HO is in a squeezed state. 
We compute explicitly  the corresponding squeezing parameter (SP) relative to the initial state at an arbitrary instant  and show that, surprisingly,  it exhibits oscillations after the first frequency jump (from $\omega_0$ to $\omega_1$), remaining constant after the second jump  (from $\omega_1$ back to $\omega_0$). We also compute the time evolution of the variance of a quadrature. 
Last, but not least, we calculate the  vacuum (fundamental state) persistence probability amplitude of the HO, as well as its  transition probability amplitude for any excited state.

%{\color{blue} We consider a harmonic oscillator (HO) with a time dependent frequency which undergoes two successive abrupt changes. By assumption, the HO starts in its fundamental state with frequency $\omega_0$, then, at $t=0$, its frequency suddenly increases to $\omega_1$ and, after a finite time interval $\tau$, it comes back to its original value $\omega_0$. Contrary to what one could naively think, this problem is a quite non-trivial one. Using algebraic methods we obtain its exact analytical solution and show that at any time $t>0$  the HO is in a squeezed state. 
%We compute explicitly  the corresponding squeezing parameter (SP) relative to the initial state at an arbitrary instant  and show that, surprisingly,  it exhibits oscillations after the first frequency jump (from $\omega_0$ to $\omega_1$), remaining constant after the second jump  (from $\omega_1$ back to $\omega_0$). We also compute the time evolution of the variance of a quadrature. 
%Last, but not least, we calculate the  vacuum (fundamental state) persistence probability amplitude of the HO, as well as its  transition probability amplitude for any excited state. }


\end{abstract}

\maketitle

\pagebreak 

\section{Introduction}\label{intro}

The classical harmonic oscillator (HO) as well as its quantum counterpart are two of the most important systems in physics \cite{Sakuray-Book, Griffiths-Book}. Their relevance relies on the ubiquity of phenomena that can be modeled by them. As a remarkable example, in quantum electrodynamics (QED) quantum harmonic oscillators are the paradigm to describe the free electromagnetic  field, hence being 
a cornerstone in quantum optics \cite{Scully-Book}. In fact, the description of a free bosonic field is frequently done by considering it as a set of HO's \cite{Greiner-Field-Book, Johnson-2002}. 
The usual quantization of the HO leads directly to the so-called Fock states, which are eigenstates of the HO hamiltonian. In quantum optics language, these Fock states correspond to $n$-photon states. Fock states of the HO are quite non-classical states, as can be seen, for instance, if we take the  quantum expectation value  of the position operator in the Heisenberg picture (or momentum operator, or even any quadrature operator) in any Fock state, which  is always zero. This is evidently  in contrast to the oscillating behaviour of the position of  a classical  HO. However, these states are very helpful and can be used as a convenient basis for describing other important states.

In the realm of QED, which describes the radiation-matter interaction with unprecedented precision,  and non-linear optics, many phenomena can be modeled by a driven time-dependent HO.
In fact, it can be shown that a HO, initially in its fundamental state which is acted by an external time dependent force, is necessarily brought into a coherent state \cite{Bo-sture-1985, Holstein-1985, Gazeau-2009, Philbin-2014, Vyas-2018}. The counterpart of this result in quantum optics is the fact that any classical current coupled to the radiation field gives rise to a coherent state of the (quantized) electromagnetic field \cite{Mandel-Wolf-book}. These states are very useful since they serve to model LASER propagation \cite{Barnett-Book-1997}. Among the main properties of coherent states we list: (\textit{i}) they saturate the Heisenberg relation (satisfy the lower-bound of the uncertainty relation); (\textit{ii}) they are eigenstates of the annihilation operator and (\textit{iii}) when the quantum expectation value of a quadrature operator is calculated in these states, the classical oscillatory behaviour is recovered.  
Notice that coherent states distribute equally the uncertainty into the quadratures.

Besides Fock states and coherent states there are many other states of the HO (and, consequently, quantum states of light) that deserve being studied, particularly, the so-called squeezed states of the HO \cite{Scully-Book, Walls-1983}. 
These states arise naturally when the parameters of a HO, namely, their mass or frequency, become time-dependent \cite{Janszky-1986, Rhodes-1989, C.F.LO-1990, Gersch-1992}. 
Squeezed states have received big attention in the last 70 years, specially because their remarkable property of having the variance of one quadrature smaller than the value associated to coherent states \textit{i.e}., the variance is squeezed, which justifies their name. Since the uncertainty relation has to be satisfied, the conjugate quadrature must have a greater uncertainty than for coherent states. It can be shown that any squeezed state can be written as a superposition of all even Fock states. As a consequence, the quantum expectation value of any quadrature operator in a squeezed state is zero and, in this sense, these states are also quite non-classical ones (a nice discussion on these states can be found in Ref. \cite{Barnett-Book-1997}).  

 Since the variance is squeezed and the uncertainty in a quadrature is closely related to a quantum limit of reduction of noise in a signal,  one of the first applications of squeezed states was in communication \cite{Yuen-1978,Yuen-1980}, as one can send information through a quadrature with reduced noise. 
Another famous application of squeezed states is in the Laser Interferometer Gravitational-Wave Observatory (LIGO) \cite{LIGO1-1992}. 
This interferometer has been conceived to measure  tiny deviations of each of its arms due to the passage of a gravitational wave. 
The first versions of such interferometers had the problem that the signal of the gravitational wave was surpassed by the zero-point fluctuations in the detector. 
One of the performed improvements consisted in using squeezed states to enhance the sensitivity of the detector by reducing the noise to signal \cite{LIGO2-2013}. 
In fact, after this improvement was made the gravitational waves were finally detected \cite{LIGO3-2016}.

There are various interesting features about time-dependent harmonic oscillators (TDHO) in addition to squeezed states. 
To mention just a few, they can be used to model the dynamical Casimir effect by using analog models in circuit QED. In fact, a double superconducting quantum interference device can be regarded, in first approximation, as a TDHO \cite{Fujii-2011}. Also the TDHO describes the quantum motion of particles in a Paul trap \cite{Kumar-1991}.
Another interesting feature of a TDHO is its connection with cosmological particle creation by non-stationary gravitational backgrounds \cite{Lemos-1987}. In order to determine the field modes in a given gravitational background, as for instance in a universe whose evolution is governed by a Freedmann-Robertson-Walker metric, one needs to solve an equation totally analogous to that of a classic harmonic oscillator with time-dependent frequency \cite{Fabio-2007,Parker-Book-2009}. In his seminal paper of 1953, Husimi showed  that the quantum solution for the TDHO can be obtained  from the corresponding classical solution \cite{Husimi-1953}.

Since Husimi's paper \cite{Husimi-1953},  a lot of progress has been achieved in the study of the TDHO. One the one hand, important  contributions were made by 
Lewis and Riesenfeld \cite{LEWIS-1969}, Popov and Perelomov \cite{Popov-1969}, 
and Malkin,  Man'ko and collaborators \cite{Malkin-1970, Man'Ko-1970} (an extensive list of references in this subject can be found in Ref.[\cite{DODONOV-2005}] and references therein).
In these papers the authors introduced the use of invariants for time-dependent hamiltonians, 
a method still very used nowadays \cite{Pedrosa-1997, Pedrosa.I.A.-1997, Andrews-1999, MOYA-2003}, sticking out its uses in shortcuts to adiabaticity \cite{Del-Campo-2011, Torrontegui-2013, Odelin-2019}. 
On the other hand, algebraic solutions for the driven TDHO have been known since the 80's,
see for instance the papers by Ma and Rhodes \cite{Rhodes-1989}, with further contributions of Lo \cite{C.F.LO-1990}. 
In these papers it is shown that the time evolution operator (TEO) at any instant can be expressed as a product of 
a squeezing, a Glauber (displacement) and a rotation operators, apart from an overall phase factor. 
Other authors  have also tackled the TDHO with distinct purposes,  from situations where a time-dependent mass was considered 
\cite{CHENG-1988, GERRY-1990, CFLO-1990,  Kumar-1991, Twamley-1993, Janszky-1994, Pedrosa-1997, Pedrosa.I.A.-1997} 
to those ones where driven forces and damping terms were included assuming different initial states for the HO 
\cite{DODONOV-1979, ABDALLA-1985, CFLO-1991, Lima-2008}. Some particular cases with exact known solutions, namely the sudden and linear frequency modulations have also been considered \cite{Janszky-1986, Kumar-1991, Janszky-1992, JANSZKY-TE-1994, Aslangul-1995, MOYA-2003}. 
Propagators of the TDHO and particular cases have also been calculated \cite{Natividade-1988, Holstein-1989, Farina-1993}
 
Although the problem of a TDHO has already been extensively investigated, there are some aspects and subtle points involving squeezing states of the HO that have not been explored in a didactic way that can be very useful for undergraduate and graduate students in the interpretation and deep understanding of the whole underlying theory. 
With the purpose of  discussing these subtle points, as well as making popular algebraic methods, we consider here a non-trivial problem of a HO with time-dependent frequency (HOTDF) which allows an exact analytical solution, namely, a HO with a frequency which undergoes two successive abrupt changes. By assumption, the HO starts in its fundamental state with frequency $\omega_0$, then, at $t=0$, its frequency suddenly increases to $\omega_1$ and, after a finite time interval $\tau$, it comes back to its original value $\omega_0$. Using algebraic methods 
based on BCH-like relations of the $su(1,1)$ Lie algebra, we obtain its exact analytical solution and show that at any time $t>0$  the HO is in a squeezed state. 
We compute explicitly  the corresponding squeezing parameter (SP) relative to the initial state at an arbitrary instant  and show that it exhibits oscillations after the first frequency jump (from $\omega_0$ to $\omega_1$), remaining constant after the second jump  (from $\omega_1$ back to $\omega_0$). We also compute the time evolution of the variance of a quadrature. 
Last, but not least, we calculate the  vacuum (fundamental state) persistence probability amplitude of the HO, as well as its  transition probability amplitude for any excited state. 
We hope this paper will approach and motivate undergraduate  and graduate  students to the increasingly important squeezing states as well as algebraic methods which  nowadays are essential in the aforesaid fields of physics. 

This paper is organized as follows: in section \ref{IntrotoSS} we present a brief introduction to the main features of squeezed states. 
In section \ref{2jumps} we  solve exactly the problem of a HOTDF where the frequency suffers two successive abrupt changes as described before. Our main results are presented in this section. Section \ref{conclusions} is left for final remarks and conclusions. For pedagogical reasons, we included an appendix \ref{app} with a brief derivation of an important BCH formula used in the text.

%%%%%%%%%%%%%%%%%%%%%%%%%%%%%%%%%%%%%%%%%%%%%%%%%%%%%%%%%%%%%%%%%%%%%%%%%%%%%%%%%%%%%%%%%%%%%%%%%%%%%%%%%%%%%%%%%%%%%%%%%%%%%%%%%%%%%%%%%%%%%%%%%%%%%%%%%%%%%%%%%%%%%%%%%%%%%%%%%%%%%%%%%%%%%%%%%%

\section{Squeezed states: main features} \label{IntrotoSS}

In this section we shall briefly quote the main features of squeezed states which will be of fundamental importance in our solution of the HOTDF. For a more complete description of squeezed states  we suggest Ref.[ \cite{Barnett-Book-1997}]. Let us consider  a HO of unit mass and constant frequency $\omega_0$, so that its hamiltonian is given by 
${\hat H}_0 = \frac{1}{2}\left({\hat p}^2 + \omega_0^2{\hat q}^2\right)$. 
Introducing the creation and annihilation operators $\hat{a}^{\dagger}$ and $\hat{a}$ in the usual way,  where $\left[\hat{a},\hat{a}^{\dagger}\right]=1$ and 
$\left[\hat{a}^{\dagger},\hat{a}^{\dagger}\right] = \left[\hat{a},\hat{a}\right] = 0 $, the previous hamiltonian can be written as 
${\hat H}_0  = (\hat n + \frac{1}{2})\omega_0$ (we are assuming $\hbar=1$), where we defined the number operator  $\hat{n} = \hat{a}^{\dagger}\hat{a}$. It can be shown that 
the corresponding energy eigenstates, denoted by $\vert n\rangle$, satisfy the eigenvalue equation 
${\hat H}_0 \vert n\rangle = (n + \frac{1}{2})\omega_0 \vert n\rangle$, where $n=0, 1, 2, ...$. 

 A squeezed state $\vert z\rangle$ is produced by the application of the squeezing operator ${\hat S}(z)$ on the fundamental state, $\vert z\rangle = {\hat S}(z) \vert 0\rangle$, with ${\hat S}(z)$ given by
\begin{equation}
\label{gensqop}
\hat{S}(z) := \exp \left\{-\frac{z}{2}\left.\hat{a}^{\dagger}\right.^{2}+\frac{z^{*}}{2}\hat{a}^{2}\right\} \, ,
\end{equation}
where $z$ is a complex number that can be conveniently written in its polar form as $z = r e^{i\varphi}$. The application of the squeezing operator written as in the previous equation on the fundamental state is not so direct due to the commutation rules between operators  $\left.\hat{a}^{\dagger}\right.^{2}$ and $\hat{a}^{2}$. However,  using well known BCH-like relations,  $\hat S(z)$ can be cast into the form \cite{Barnett-Book-1997}
\begin{eqnarray}
\label{gensqopbch}
\hat{S}(z)&=&\exp \left\{-\frac{1}{2} \left.\hat{a}^{\dagger}\right.^{2} \exp{(i\varphi)}\tanh(r)\right\}\exp \left\{-\frac{1}{2}(\hat{a}^{\dagger}\hat{a}+\hat{a}\hat{a}^{\dagger})\ln{(\cosh(r))} \right\} \times\nonumber\\
&& \times\exp \left\{\frac{1}{2}\hat{a}^{2} \exp{(-i\varphi)}\tanh(r)\right\} \, .
\end{eqnarray}
Now, recalling that $\hat a\vert 0\rangle = 0$, $(\hat{a}^{\dagger}\hat a + \hat a\hat{a}^{\dagger}) \vert n\rangle = (2n + 1)\vert n\rangle$ and 
$\hat{a}^{\dagger}\vert n\rangle = \sqrt{n+1}\vert n+1 \rangle$, it is straightforward to show that  \cite{Barnett-Book-1997}
%
%shows that a squeezed state, in the Fock basis, is a superposition solely of the even states  \cite{Barnett-Book-1997}
%
\begin{equation}
\left|z\right\rangle = {\hat S}(z) \vert 0\rangle = \sqrt{\mbox{sech}(r)}\sum_{n=0}^{\infty}\frac{\sqrt{(2n)!}}{n!}
\left[-\frac{1}{2}e^{(i\varphi)}\tanh(r)\right]^{n}\left|2n\right\rangle.
\label{eq:squeezed}
\end{equation}
As evident from the previous equation, a squeezed state is a superposition solely of the even number states and therefore they are quite non-classical states. Note, for instance, that as it also occurs with Fock states 
$\{\vert n\rangle\}$, the expectation value of momentum or position operators in a squeezed state is zero, a result far from an expected classical behaviour.

\begin{figure}[h!]
\centering
\includegraphics[width=0.65\linewidth]{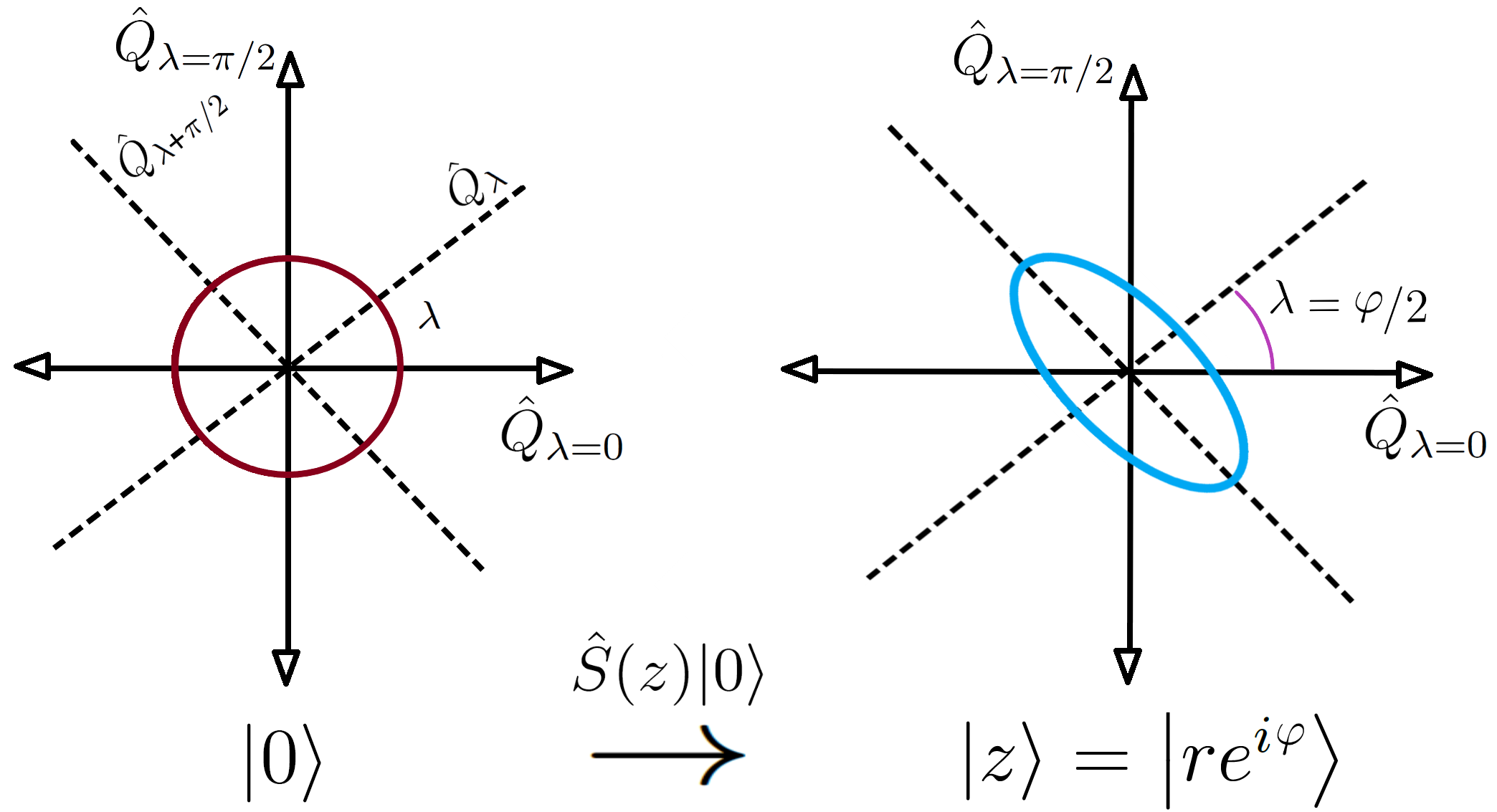}
\caption{Schematic diagram of the uncertainty areas in the generalized 
%coordinate and momentum (X, P) 
quadrature phase space of a squeezed state (oval) and the fundamental state (circle).}
\label{fig:charuto}
\end{figure}

As can be inferred from Eq. (\ref{eq:squeezed}), $r$ and $\varphi$ determine uniquely the squeezed state. In order to interpret them, it is convenient to introduce the  quadrature operator ${\hat Q}_\lambda$, defined by  \cite{Barnett-Book-1997}
\begin{equation}
\hat{Q}_{\lambda}=\frac{1}{\sqrt{2}}\left[ \hat{a}^{\dagger}e^{i\lambda} + \hat{a}e^{-i\lambda} \right]\, ,
\label{eq:quadratureop}
\end{equation}
which satisfies the commutation relation $[{\hat Q}_\lambda,{\hat Q}_{\lambda + \pi/2}] = i \; $ for any real $\lambda$. It is evident from the previous definition that 
$\hat{Q}_{\lambda=0} = (\hat{a}^{\dagger} +  \hat{a})/\sqrt{2}= \sqrt{\omega_{0}}\hat{q} $, so that $\hat{Q}_{\lambda=0}$ and $ \hat{q}$ are proportional, and that 
$\hat{Q}_{\lambda=\pi/2} = i(\hat{a}^{\dagger} -  \hat{a})/\sqrt{2}=\frac{\hat{p}}{\sqrt{\omega_{0}}}$, so that $\hat{Q}_{\lambda=\pi/2}$ and $\hat{p}$ are proportional. 
We say that the HO is in a squeezed state (or simply squeezed) if the variance of one of the quadrature operators in this state is smaller than $\frac{1}{2}$.  
It can be shown that the variance of  the quadrature operator ${\hat Q}_\lambda$ in the generic squeezed state $\vert z\rangle$ written in Eq. (\ref{eq:squeezed}) is given by  \cite{Barnett-Book-1997}
\begin{eqnarray}
 \left(\Delta Q_\lambda \right)^2 &=& \langle z\vert\hat{Q}_{\lambda}^{2}\vert z\rangle - (\langle z\vert \hat{Q}_{\lambda}\vert z\rangle)^2\cr
&=&
\frac{1}{2}\left[e^{2r} \sin^{2}\left(\lambda-\varphi/2\right)+e^{-2r} \cos^{2}\left(\lambda-\varphi/2\right)\right]\, .
\label{eq:quadratureopvariance}
\end{eqnarray}
Note the explicit dependence of $\left(\Delta Q_\lambda\right)^2$ with $r$ and $\varphi$. Further, a direct inspection of the previous equation shows that 
$ \left(\Delta Q_\lambda \right)^2 $ must satisfy the inequalities
\begin{equation}
 \frac{e^{-2r}}{2} \le  \left(\Delta Q_\lambda \right)^2 \le  \frac{e^{2r}}{2}\, ,
\end{equation}
which justifies the interpretation of  $r$ as the squeezing parameter (SP). Parameter $\varphi$ is referred to as the squeezing phase (SPh). 
Note also that, the variances of a quadrature operator in a given squeezed state depends on $\lambda$, in contrast to what happens with coherent states (the variances of all quadrature operators are the same for coherent states). 
Furthermore, a given quadrature operator, for instance, operator ${\hat Q}_{\lambda=0}$, may have different variances for different squeezed states even if these states have the same SP. 
It suffices that these states have different values of the SPh. All of the above results are graphically explained in Figure 
\ref{fig:charuto}. In these graphics, the variances of a quadrature operator for a given $\lambda$ can be visualized as the distance between the two intersecting points of the circle (fundamental state) and the oval (squeezed state) with the straight line characterizing the quadrature operator under consideration. It can be shown that, for a squeezed state given by $\ket{r e^{i \varphi}}$, the quadrature operator that has the minimum variance is the one with $\lambda = \varphi/2$, as indicated in Fig.\ref{fig:charuto}.
%
%%%%%%%%%%%%%%%%%%%%%%%%%%%%%%%%%%%%%%%%%%%%%%%%%%%%%%%%%%%%%%%%%%%%%%%%%%%%%%%%%%%%%%%%%%%%%%%%%

We finish this section by studying the time evolution of a squeezed state of the hamiltonian $\hat{H}_{0}$ caused by the action of the TEO associated to the same hamiltonian, that is
\begin{eqnarray}
e^{-i{\hat H}_0 t}\left|z\right\rangle &=& \sqrt{\mbox{sech}(r)}\sum_{n=0}^{\infty}\frac{\sqrt{(2n)!}}{n!}
\left[-\frac{1}{2}e^{(i\varphi)}\tanh(r)\right]^{n}\, e^{-i{\hat H}_0 t}\,\left|2n\right\rangle \cr
&=&
e^{-\frac{i}{2}\omega_0 t}\, \sqrt{\mbox{sech}(r)}\sum_{n=0}^{\infty}\frac{\sqrt{(2n)!}}{n!}
\left[-\frac{1}{2}e^{i(\varphi - 2\omega_0 t)}\tanh(r)\right]^{n}\,\left|2n\right\rangle \cr
&=& \vert z e^{-2i\omega_0 t}\rangle \, ,
\label{eq:ssevol}
\end{eqnarray}
where we have neglected the overall phase coming from the zero-point energy. As it can be noted, the time evolution introduces dynamics in the squeezed state only through the SPh. Graphically, this is equivalent to the spinning of the oval of Figure \ref{fig:charuto} with a period of $\pi/\omega_{0}$, since now
\begin{equation}
 \left(\Delta Q_\lambda \right)^2 (t) = 
\frac{1}{2}\left[e^{2r} \sin^{2}\left(\lambda-\varphi/2 +\omega_{0}t\right)+e^{-2r} \cos^{2}\left(\lambda-\varphi/2 +\omega_{0}t \right)\right]\, .
\label{eq:quadratureopvariance2}
\end{equation}
%

\begin{comment}
For future convenience (in order to make comparisons with results in the literature), in the calculations of this section we will use the particular definition of 
quadrature (scaled) operators introduced by Janszky \cite{Janszky-1994}. They are given by $1/\sqrt{2}$ times the quadrature operator defined in Eq. (\ref{eq:quadratureop}). For instance, for particular values of $\lambda$ we have: $\hat{X}:=(\frac{1}{\sqrt{2}})\hat{Q}_{\lambda=0}$ and $\hat{P}:=(\frac{1}{\sqrt{2}})\hat{Q}_{\lambda=\pi/2}$. As a consequence, squeezing will occur when the variance is smaller than $1/4$ instead of $1/2$. In fact,  the scaled operators $\hat{X}$ and $\hat{P}$ can be identified with the real and imaginary parts of $\hat{a}$ and $\hat{a}^{\dagger}$: $\hat{a}=\hat{X}+i\hat{P}$ and $\hat{a}^{\dagger}=\hat{X}-i\hat{P}$ \cite{Janszky-1994}.   
\end{comment}

%%%%%%%%%%%%%%%%%%%%%%%%%%%

%%%%%%%%%%%%%%%%%%%%%%%%%%%

\section{Harmonic oscillator with two successive frequency jumps} \label{2jumps}

\begin{figure}[h!]
\centering
\includegraphics[width=.38 \linewidth]{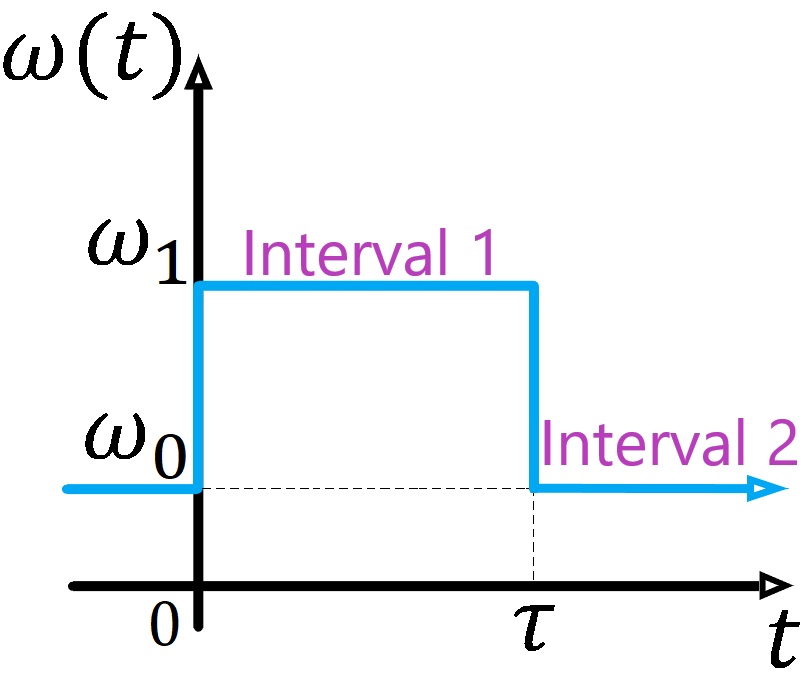}
\caption{Frequency $\omega(t)$ given by Eq. (\ref{Omega(t)}) as a function of time.}
\label{fig:frequpdown}
\end{figure}

In this section, we present the exact solution of a HO that undergoes two successive abrupt changes in its frequency. Initially, the frequency jumps from $\omega_{0}$ to $\omega_{1}$ at $t=0$. From $t=0$ until $t=\tau$ (interval 1) the frequency remains constant with value $\omega_1$. Then, at $t=\tau$,  the HO frequency jumps back to its original value $\omega_{0}$,  remaining constant for $t>\tau$ (interval 2). Therefore, $\omega(t)$ can be written as
\begin{equation}
\label{Omega(t)}
\omega(t) = \omega_0 + (\omega_1 - \omega_0)\left[\Theta(t) - \Theta(t - \tau)\right]\, ,
\end{equation}
where $\Theta$ is the usual Heaviside step function. Figure \ref{fig:frequpdown} shows $\omega(t)$ as a function of time. 

In order to compute the quantum state of the HO at an arbitrary instant $t>0$, we need to obtain the time evolution operator for this time dependent hamiltonian. This will be done explicitly 
 with the aid of algebraic methods in the next subsections.

%We shall show that in interval 1, just after the first frequency change, the state of the system become squeezed with their parameters time-dependent. Then, in interval 2, immediately after the %second frequency change, the SP becomes fixed in time and all the dynamics of the state is given by the SPh.

\subsection{Time evolution operator}

For convenience, let us write the hamiltonian of the HO with a time-dependent frequency given by Eq. (\ref{Omega(t)}) in the form
\begin{equation}
\hat{H}=\frac{\hat{p}^{2}}{2}+\frac{1}{2}\Bigl(\omega_{0}^{2}+2\omega_{0}\eta(t)\Bigr)\hat{q}^{2} \, ,
\end{equation}
so that the time dependence of the frequency is encoded in  $\eta(t)$, which couples with the square of the position operator and is defined by
\begin{equation}
\eta(t) = \eta_0\Bigl\{ \Theta(t) - \Theta(t - \tau)\Bigr\}\, ,
\end{equation}
where 
\begin{equation}\label{eq:etafunc}
\eta_0 = \frac{\left.\omega_{1}\right.^{2}-\left.\omega_{0}\right.^{2}}{2\omega_0} \;\;\;\; \Longrightarrow\;\;\;\; \omega_{1} = \sqrt{\omega_{0}^{2}+2\omega_{0}\eta_{0}}\, .
\end{equation}
Previous equations mean that for the aforementioned time-intervals 1 and 2, the corresponding time-independent hamiltonian operators, $\hat H_1$ and $\hat H_2$, which describes the time evolution of the HO, are given respectively by
\begin{equation}\label{H1-2}
\hat H_1 = \frac{\hat{p}^{2}}{2} + \frac{1}{2}\omega_{1}^{2}\hat{q}^{2} \;\;\;\;\mbox{and}\;\;\;\;
\hat H_2 = \frac{\hat{p}^{2}}{2} + \frac{1}{2}\omega_{0}^{2}\hat{q}^{2}\, .
\end{equation}
Consider the initial state as the fundamental state, $\left| \psi(t=0) \right\rangle=\left| 0 \right\rangle$, so that the state  of the HO at any instant $ t > 0 $ is given by 
$\left| \psi(t) \right\rangle=\hat{U}(t,0)\left| 0 \right\rangle$. Since the frequency on each interval is different, we consider separately the TEO in each interval. For any $t$ belonging to interval 1, namely, $0< t \leq \tau$, we have 
\begin{equation}
\left| \psi(t) \right\rangle = \hat{U}_1(t,0)\left| 0 \right\rangle =  e^{-i\hat H_1 t} \left| 0 \right\rangle\, ,
\label{eq:wavefunc}
\end{equation}
while for any $t$ belonging to interval 2, namely, $t > \tau$, we can use the composition property of the TEO to write  
\begin{equation}
\left| \psi(t) \right\rangle=\hat{U}(t,0) \left| 0 \right\rangle = \hat{U}_{2}(t,\tau)\hat{U}_{1}(\tau,0)\left| 0 \right\rangle 
= e^{-i\hat H_2 (t - \tau)}  e^{-i\hat H_1 \tau}   \left| 0 \right\rangle\, .
\label{eq:TEOcomp}
\end{equation}
In terms of operators $\hat a$ and $\hat{a}^{\dagger}$ the TEO's  are written as
\begin{eqnarray}
\hat{U}_{1}(t,0) &=&  exp\left(-i\hat{H}_{1}t\right) = 
\exp\left\{-\frac{i}{2}\biggl( \left( \hat{a}^{\dagger^{2}}+\hat{a}^{2}\right) \eta_{0}+\left(\hat{a}^{\dagger}\hat{a}+\hat{a}\hat{a}^{\dagger}\right) \left(\omega_{0}+\eta_{0}\right)\biggl)t \right\} \\
\hat{U}_{2}(t,\tau)
&=& \exp\left(-i\hat{H}_{2}(t-\tau)\right) = 
\exp\left\{-\frac{i}{2}\left(\hat{a}^{\dagger}\hat{a}+\hat{a}\hat{a}^{\dagger}\right)\omega_{0}(t-\tau) \right\}. 
\label{eq:TEOSann}
\end{eqnarray}
To obtain the final state $\left|\psi(t)\right\rangle = \hat{U}_{2}(t,\tau)\hat{U}_{1}(\tau,0)\left|0 \right\rangle$ is not a simple task, since the exponents of both TEO's contain non-commuting operators. However, whenever a hamiltonian is written as a linear combination of  the generators of a given Lie algebra, it is possible to factorize the TEO writing it as a product of exponentials containing each one only one generator of the corresponding Lie algebra \cite{CFLO-1990}. Explicitly, if $\hat A_1$, $\hat A_2$, ..., $\hat A_N$ are the $N$ generators of a given Lie algebra, they will satisfy
\begin{equation}
\left[\hat{A}_{i},\hat{A}_{j}\right]=\sum_{k=1}^{k=N}C_{i j}^{k}\hat{A}_{k}\:\:\:\:\:\: \mbox{for $j\neq i$}\, ,
\label{eq:Lie}
\end{equation}
where coefficients $C_{i j}^{k}$ are called structure constants of the corresponding Lie group. For a connected Lie group, any element of the group, say $g$, can be written, for instance, in the two following ways:
\begin{equation}
g = \exp\left(\theta_{1}\hat{A}_{1}+\theta_{2}\hat{A}_{2}+...+\theta_{N}\hat{A}_{N}\right) \:\:\:\:\:
\mbox{or}
\:\:\:\:\:
g = \exp\left(f_{1}\hat{A}_{1}\right)\exp\left(f_{2}\hat{A}_{2}\right)...\exp\left(f_{N}\hat{A}_{N}\right).
\label{eq:BCH}
\end{equation}
Of course parameters $f_1$, $f_2$, ..., $f_N$ will be different from parameters $\theta_{1}$, $\theta_{2}$, ..., $\theta_{N}$. However, from the commutation relations of the Lie algebra it is possible to find the algebraic relations between these two sets of parameters, $f_{1}(\theta_{1},\theta_{2},...,\theta_{N})$, $f_{2}(\theta_{1},\theta_{2},...,\theta_{N})$,...,$f_{N}(\theta_{1},\theta_{2},...,\theta_{N})$. These relations are referred to as BCH-like formulas of the Lie algebra (BCH comes from the original papers by Baker, Campbell and Hausdorff \cite{Baker-1901,Baker-1905,Campbell-1896,Campbell-1897,Hausdorff-1906}).

In order to solve the problem under consideration our strategy will be to use the different algebraic representations given in Eq. (\ref{eq:BCH}), since, as we will show, the hamiltonian of our problem can be written as a linear combination of the generators of a known Lie algebra. With this goal in mind, we first identify three important operators present in the argument of the TEO's, namely, $\hat{a}^{\dagger^{2}}$, $\hat{a}^{2}$  and  $\hat{a}^{\dagger}\hat{a}+\hat{a}\hat{a}^{\dagger}$. Using the following definitions:
\begin{equation}
\hat{K}_{+} := \frac{\hat{a}^{\dagger^{2}}}{2}, \:\:\: \hat{K}_{-} := \frac{\hat{a}^{2}}{2} \:\:\:  \mbox{and} 
\:\:\:\hat{K}_{c} := \frac{\hat{a}^{\dagger}\hat{a}+\hat{a}\hat{a}^{\dagger}}{4} = \frac{1}{2}\left( {\hat a}^\dagger \hat a + \frac{1}{2}\right),\
\label{eq:LieAlgebraGen}
\end{equation}
it is straightforward to show that:
\begin{equation}
\left[\hat{K}_{-},\hat{K}_{+}\right]=2\hat{K}_{c}, \:\:\: \:\:\: \left[\hat{K}_{c},\hat{K}_{\pm}\right]=\pm\hat{K}_{\pm}\, ,
\label{eq:algebraK}
\end{equation}
which are identified as the commutation relations of the $\textit{su}(1,1)$ Lie algebra \cite{Barnett-Book-1997, Truax-1985}. 
Hence, the TEO's can be written as:
\begin{equation}
\hat{U}_{1}(t,0) = e^{\lambda_{+}(t)\hat{K}_{+} + \lambda_{-}(t)\hat{K}_{-} + \lambda_{3}(t)\hat{K}_{c}}, 
\:\:\:\:\:\:\hat{U}_{2}(t,\tau) = e^{\lambda_{4}(t)\hat{K}_{c}},
\label{eq:TEOSfinal}
\end{equation}
where
\begin{equation}
\lambda_{+}(t) = \lambda_{-}(t) = -i\eta_{0}t, \;\;\;\; 
\lambda_{3}(t) = -2i(\omega_{0} + \eta_{0})t,\;\;\;\;
\lambda_{4}(t) = -2i\omega_{0}(t-\tau).
\label{eq:lambdaspeques}
\end{equation}
Using an appropriate BCH formula for the TEO $\hat{U}_{1}(t,0)$, it can be factorized in the following way \cite{ Truax-1985} (see Appendix A for a brief demonstration) 
\begin{equation}
\hat{U}_{1}(t,0) = e^{\lambda_{+}(t)\hat{K}_{+} + \lambda_{-}(t)\hat{K}_{-} + \lambda_{3}(t)\hat{K}_{c}} 
 = e^{\Lambda_{+}(t)\hat{K}_{+}} e^{(\ln\Lambda_{3}(t))\hat{K}_{c}} e^{\Lambda_{-}(t)\hat{K}_{-}},
\label{eq:BCH1}
\end{equation}
where
\begin{equation}
%\label{true4}
\Lambda_{3} = \left(\cosh(\nu)-\frac{\lambda_{3}}{2\nu}\sinh(\nu)\right)^{-2} \;\;\; \mbox{and} \;\;\;
\label{true5}
\Lambda_{\pm} = \frac{2\lambda_{\pm}\sinh(\nu)}{2\nu \cosh(\nu)-\lambda_{3}\sinh(\nu)}\, ,
\end{equation}
and $\nu$ is given by
\begin{equation}
\label{eq:nu}
\nu^{2} = \frac{1}{4}\lambda_{3}^{2}-\lambda_{+}\lambda_{-} \,\, .
\end{equation}
In order to simplify the notation,  we have omitted the (time dependent) arguments of the parameters. From now on, we shall do that whenever there is no risk 
 of confusion. 
 
From Eq. (\ref{eq:BCH1})  we see that the  state  of the HO at any instant of interval 1 can be written as
\begin{equation}
\left| \psi(t) \right\rangle = \hat{U}_{1}(t,0)  \vert 0\rangle = e^{\Lambda_{+}\hat{K}_{+}}e^{(\ln\Lambda_{3})\hat{K}_{c}}e^{\Lambda_{-}\hat{K}_{-}}\left|0\right\rangle\, .
\label{eq:WFR1}
\end{equation}
Analogously, the  state  of the HO at any instant of interval 2 can be written as
\begin{equation}
\left| \psi(t) \right\rangle =   \hat U_2(t,\tau) \vert \psi(\tau)\rangle =  e^{\lambda_{4}\hat{K}_{c}}\left| \psi(\tau) \right\rangle\, ,
\label{eq:wavefunck}
\end{equation}
where we used the second equation written in (\ref{eq:TEOSfinal})  and $\left| \psi(\tau) \right\rangle$ is obtained from Eq. (\ref{eq:WFR1}) simply by taking in this equation $t = \tau$. 
In the next subsection we shall use the specific properties of operators  $\hat{K}_{-}$, $\hat{K}_{+}$ and $\hat{K}_{c}$ to calculate the state of the HO at any instant of time and to show that it consists of a squeezed state.

%%%%%%%%%%%%%%%%%%%%%%%%%%%%%%%%%%%%%%%%%%%%%%%%%%%%%%%%%%%%%%%%%%%%%%%%%%%%%%%%%%%%%%%%%%%%%%%%%%

\subsection{Results and discussions} \label{jbf}

In order to calculate explicitly the  state of the HO at any instant, we need to know how operators  $\hat{K}_{-}$,  $\hat{K}_{+}$ and  $ \hat{K}_{c}$ act on the energy eigenstates. 
From their definitions, it is immediate to check the following relations,
\begin{eqnarray}
\hat{K}_{-}\left|n\right\rangle&=&\frac{1}{2}\sqrt{n(n-1)}\left|n-2\right\rangle, \nonumber\\
\hat{K}_{+}\left|n\right\rangle&=&\frac{1}{2}\sqrt{(n+1)(n+2)}\left|n+2\right\rangle, \\
\hat{K}_{c}\left|n\right\rangle&=&\frac{1}{2}(n+\frac{1}{2})\left|n\right\rangle\, ,\nonumber
\label{Eq:rules}
\end{eqnarray}
which will be useful in the subsequent discussion. For convenience, and also for the sake of clarity,  we shall present the results for the time intervals 1 and 2 separately.

\centerline{\bf Interval 1 }

Let us compute here the state of the  HO at an arbitrary  instant $t$ such that $0 < t \le \tau$. 
 Using previous equations, it can be shown that
\begin{equation}
e^{\Lambda_{-}\hat{K}_{-}}\left|0\right\rangle=\left|0\right\rangle \;\;\;\;\;\;\;\; \mbox{and} \;\;\;\;\;\;\;\;e^{\ln(\Lambda_{3})\hat{K}_{c}}\left|0\right\rangle=\left(\Lambda_{3}\right)^{1/4}\left|0\right\rangle\, .
\label{eq:appc}
\end{equation}
Therefore Eq. (\ref{eq:WFR1}) is reduced to
\begin{equation}
\left| \psi(t) \right\rangle=\left(\Lambda_{3}\right)^{1/4}e^{\Lambda_{+}\hat{K}_{+}}\left|0\right\rangle\, .
\label{eq:WFR1prima}
\end{equation}
Note that in the last expression the coefficient $\Lambda_{-}$ (associated to operator $\hat{K}_{-}$) does not appear.  
Finally, since $\hat{K}_{+} = \frac{1}{2}\hat{a}^{\dagger^{2}}$,  the above equation takes the form
%the application of the operator $e^{\hat{K}_{+}}$ on the fundamental state leads to
%
\begin{equation}
\left|\psi(t)\right\rangle=\sqrt{\left|\Lambda_{3}\right|^{1/2}}\sum_{n=0}^{\infty}\frac{\sqrt{(2n)!}}{n!}
\left[\frac{1}{2}\left|\Lambda_{+}\right|e^{i\vartheta}\right]^{n}\left|2n\right\rangle \, ,
\label{eq:w1final}
\end{equation}
where the overall phase has been removed by writing
\begin{equation}
\Lambda_{3}=\left|\Lambda_{3}\right|e^{i\chi}, \:\:\:\: \mbox{and}\:\:\:\: \Lambda_{+}=\left|\Lambda_{+}\right|e^{i\vartheta}.
\label{eq:phasecoeff}
\end{equation}
Comparison of Eqs. (\ref{eq:w1final}) and (\ref{eq:squeezed}) strongly suggests that the state of the HO after the frequency jump $\omega_0 \rightarrow \omega_1$ is a squeezed state with respect to the original hamiltonian (describing the HO with a constant frequency $\omega_0$).  However, for this to be true we must be able to make the following identifications:  
\begin{equation}\label{Identifications}
\left|\Lambda_{3}\right|^{1/2} =  \mbox{sech}(r) \;\;\;\mbox{and}\;\;\;  \left|\Lambda_{+}\right| = \tanh(r)\, ,
\end{equation}
which implies the  relation $\left|\Lambda_{3}\right|+\left|\Lambda_{+}\right|^{2} = 1 $. Hence, all we have to do is to demonstrate this relation.  
%In fact, this condition can be straightforwardly verified, as it is shown in Apendix B.
With this purpose in mind, first note that from Eqs. (\ref{true5}), $\Lambda_{+}$ can be written as a function of $\Lambda_{3}$ as
\begin{equation}
\label{eq:lambamb}
\Lambda_{+} = \lambda_{+}\frac{\sinh(\nu)}{\nu}(\Lambda_{3})^{1/2}\, .
\end{equation}
Now, from Eqs. (\ref{eq:lambdaspeques}) and Eq. (\ref{eq:nu}) we see that $\nu$ is purely imaginary, so that $\sinh(\nu) / \nu$ is real valued. Besides, using the fact that 
$\left.\lambda_{+}\right.^{*}=-\lambda_{+} $, we can easily show that $\left|\Lambda_{+}\right|^{2}$ is given by
\begin{equation}
\label{eq:lamb+modulus}
\left|\Lambda_{+}\right|^{2} = -(\lambda_{+})^{2} \left(\frac{\sinh(\nu)}{\nu}\right)^{2} \left|\Lambda_{3}\right|  \, ,
\end{equation}
so that
\begin{equation}
\label{eq:identwo}
\left|\Lambda_{3}\right|+\left|\Lambda_{+}\right|^{2}  = 
\left[1 -(\lambda_{+})^{2}\frac{\sinh^{2}(\nu)}{\nu^{2}} \right] \left|\Lambda_{3}\right|\, .
\end{equation}
Now, it follows from Eq. (\ref{true5}) that
\begin{equation}
\label{eq:lamb3modu}
\left|\Lambda_{3}\right| = \frac{1}{\cosh^{2}(\nu) - (\frac{\lambda_{3}}{2})^{2} \left(\frac{\sinh(\nu)}{\nu}\right)^{2}}
=
\frac{1}{1- \left(\frac{\lambda_{3}^2}{4}  - \nu^{2} \right) \frac{\sinh^{2}(\nu)}{\nu^{2}}}\, ,
\end{equation}
where we used the fact  that $\lambda_3$ is purely imaginary and that $\sinh(\nu) / \nu$ is real valued, and in the last step we used the identity $\cosh^{2}{\nu} = 1 + \sinh^{2}{\nu}$. 
Substituting Eq. (\ref{eq:lamb3modu}) into Eq. (\ref{eq:identwo}), we  finally obtain
\begin{equation}
\left|\Lambda_{3}\right|+\left|\Lambda_{+}\right|^{2}  = 
\frac{1 - \lambda_{+}^{2}\frac{\sinh^{2}(\nu)}{\nu^{2}}}
{1- \left(\frac{\lambda_{3}^2}{4}  - \nu^{2} \right) \frac{\sinh^{2}(\nu)}{\nu^{2}}} =  1\, ,
\end{equation}
where  we used the relation $(\frac{\lambda_{3}}{2})^{2}  - \nu^{2} = {\lambda_{+}}^{2}$, which follows directly from Eqs. (\ref{eq:lambdaspeques}) and Eq. (\ref{eq:nu}).
Hence we conclude that a quantum HO initially in its fundamental state will evolve inevitably, after a sudden change in its frequency, into a squeezed state.

\begin{figure}[h!]
\centering
\includegraphics[width=0.6\linewidth]{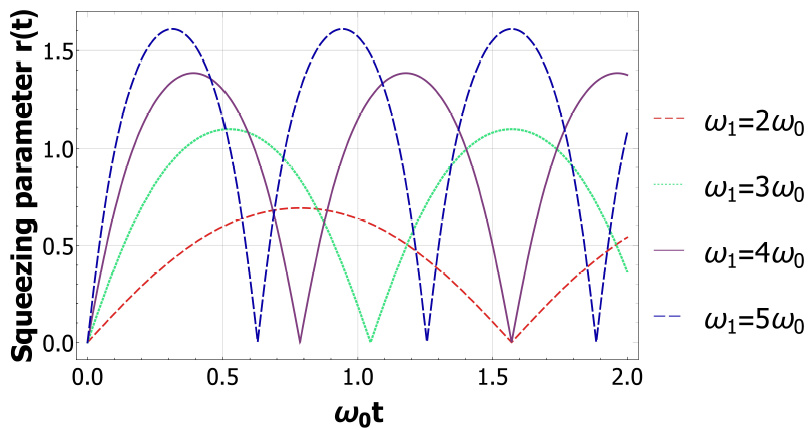}
\vskip -0.5cm
\caption{Squeezing parameter $r(t)$ given by Eq. (\ref{eq:SPfreq}) as a function of time $\omega_{0}t$ for different values of the frequency $\omega_{1}$ achieved after the first frequency jump.}
\label{rVersusTau}
\end{figure}

Let us now obtain one important result of this work, namely,  an explicit expression for the time evolution of the squeezing parameter r in interval 1 ($0 < t \le \tau$). With this purpose, from the first equation written in (\ref{Identifications}), $\mbox{sech}(r) = \left|\Lambda_{3}\right|^{1/2}$, we have
\begin{equation}\label{Idensqpar1}
%\mbox{sech}(r) = \left|\Lambda_{3}\right|^{1/2} \;\;\;\;\;\; \Longrightarrow \;\;\;\;\;\;  
r =\mbox{arcosh}\left( 1/\left|\Lambda_{3}\right|^{1/2} \right) \, .
\end{equation}
Substituting Eq. (\ref{eq:lamb3modu})  into the previous equation and using the relation $(\frac{\lambda_{3}}{2})^{2}  - \nu^{2} = {\lambda_{+}}^{2}$, we get
\begin{equation}\label{Idensqpar2}
r =\mbox{arcosh}\left( \left[1- (\lambda_{+})^{2} \frac{\sinh^{2}(\nu)}{\nu^{2}} \right]^{1/2} \right) \, .
\end{equation}
From Eqs. (\ref{eq:lambdaspeques}) it can be easily shown that $(\lambda_{+})^{2}=- \left.\eta_{0}\right.^{2} t^{2}$ and 
$\nu^{2} = -\left( \left.\omega_{0}\right.^{2} + 2 \omega_{0}\eta_{0} \right) t^{2}$. Using Eq. (\ref{eq:etafunc}) to re-express the above results as a function of the frequency $\omega_{1}$  we obtain
\begin{equation}\label{eq:nuandlambda+}
\nu^{2} = -\left.\omega_{1}\right.^{2} t^{2} \;\;\;\; \Longrightarrow \;\;\; \nu = \pm i\omega_{1} t \, , \:\:\:\:\:
(\lambda_{+})^{2} = - \left(\frac{\left.\omega_{1}\right.^{2}-\left.\omega_{0}\right.^{2}}{2\omega_{0}}\right)^{2} t^{2} \, .
\end{equation}
Finally, substituting last results into Eq. (\ref{Idensqpar2}), we obtain
%From Eqs. (\ref{eq:paritrig}) $\sinh^{2}(\pm i\omega_{1} t)=-\sin^{2}(\omega_{1} t)$ and then, substituting last results in Eq. (\ref{Idensqpar2}), we have
%
\begin{equation}
r(t) = \mbox{arcosh}{\sqrt{1+\left(\frac{\left.\omega_{1}\right.^{2}-\left.\omega_{0}\right.^{2}}{2\omega_{0}\omega_{1}}\right)^2 \sin^{2}\left(\omega_{1}t \right)}} \, ;
\;\;\;\; (0 < t \le \tau)\, .\,
\label{eq:SPfreq}
\end{equation}
where we used that $\sinh^{2}(\pm i\omega_{1} t)=-\sin^{2}(\omega_{1} t)$. 
Previous equation gives the time evolution of the SP. A direct inspection of Eq. (\ref{eq:SPfreq}) shows that $r(t)$ is a periodic function of $t$ with period equal to 
$\pi/\omega_1$, since $r(t + \pi/\omega_1) = r(t)$ for any $t$ in the interval 1. 
In Fig.\ref{rVersusTau} we plot $r(t)$ as a function of $t$, with $0 < t \le \tau$, for  a fixed value of $\omega_0$  but 
different values of $\omega_1$ (for convenience, but without loss of generality for our purposes, we will use $\omega_0 = 1$). 

 As it can be seen from Fig.\ref{rVersusTau} , $r(t)$ has indeed a periodic behaviour whose period decreases as $\omega_1$ increases, as expected.  
It is worth mentioning that, the way we defined the SP $r(t)$, it is a continuous function of $t$. For instance, no matter how big the frequency jump $\omega_0 \rightarrow \omega_1$ may be
 (but assuming always a finite $\omega_1$), $r(t)$ starts growing from zero continually. Of course, the greater the jump, the greater will be the slope of $r(t)$ near $t=0$. But in the limit $t \rightarrow 0$ we always have $r(t) \rightarrow 0$, as can be seen in Fig.\ref{rVersusTau}. The continuity of $r(t)$ is a direct consequence of the continuity of the state vector $\vert \psi(t)\rangle$, 
 \begin{equation}
 \lim_{t\rightarrow 0} \vert\psi(t)\rangle =  \lim_{t\rightarrow 0} e^{-i{\hat H}_1t}  \vert 0 \rangle = \vert 0\rangle\, .
 \end{equation}
 Also, note that the bigger the frequency jump is the higher the maximum value achieved by the SP will be, a result that can be checked by a direct inspection of Fig.\ref{rVersusTau}. 
 This maximum value of $r(t)$ is achieved for the first time at $t = \pi/(2\omega_1)$.

\begin{figure}[h!]
\begin{center}
\includegraphics[width=0.5\linewidth]{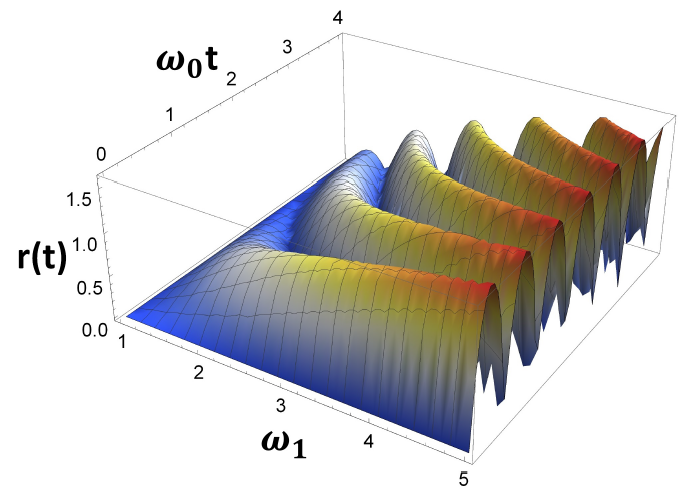}
\end{center}
\vskip -0.5cm
\caption{Tridimensional plot of $r(t)$ as a function of time $\omega_{0}t$ and $\omega_1$. Intersections  with vertical planes with constant $\omega_1$ give  $r(t)$ as a function of $\omega_{0}t$ for different values of $\omega_1$. Analogously,  Intersections with vertical planes with constant $t$ give $r(t)$ as a function of $\omega_1$ for different values of $\omega_{0}t$.}
 \label{rVersusEtaetau}
\end{figure}

We could now analyse how $r(t)$ at an arbitrary (fixed) instant  behaves as we vary $\omega_1$.  This behaviour will depend on the characteristics of the function $\mbox{arcosh}$, as well as on its argument. 
However, although an oscillatory behaviour will appear, due to the presence of $\sin^2(\omega_1 t)$ in Eq. (\ref{eq:SPfreq}),  $r(t)$ will not be a periodic function of $\omega_1$ anymore, since the argument of 
$\mbox{arcosh}$ is not a periodic function of $\omega_1$.  All these features can be appreciated in the \textit{3D} plot of Fig.\ref{rVersusEtaetau}. 
In this figure, on one hand the behaviour of $r(t)$ as a function of $\omega_1$ (with fixed $t$) can be obtained by intersecting it with vertical planes of constant values of $t$. On the other hand, if we intersect Fig.\ref{rVersusEtaetau} with vertical planes of constant values of $\omega_1$ we reobtain the plots shown in Fig.\ref{rVersusTau}.

\begin{figure}[h!]
\begin{center}
\includegraphics[width=0.65 \linewidth]{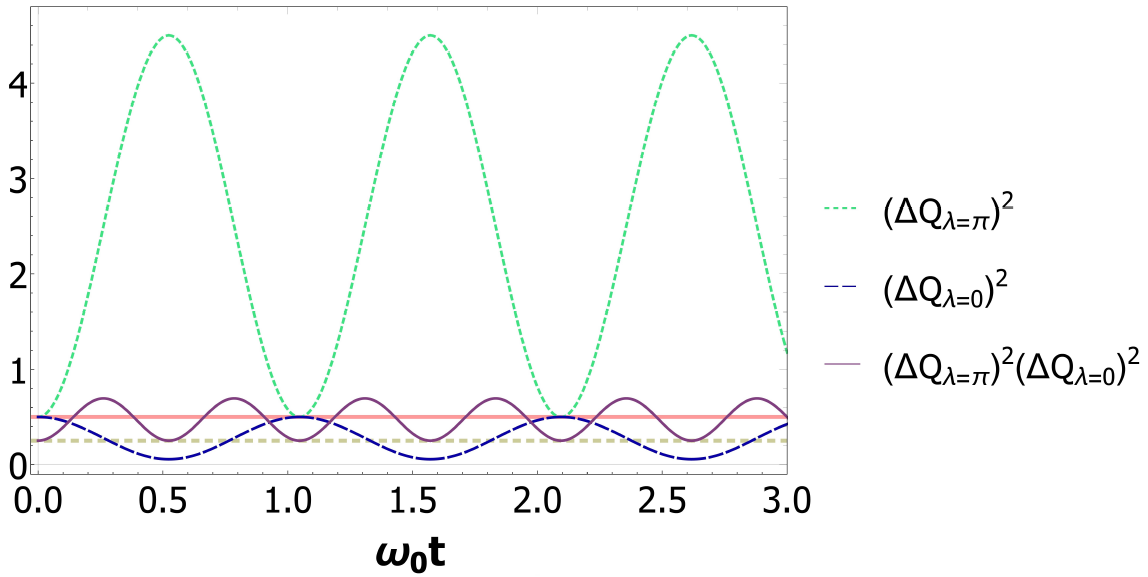}
\end{center}
\vskip -1.0cm
\caption{Variances $(\Delta Q_{\lambda=0})^2$ and $(\Delta Q_{\lambda=\pi/2})^2$ as functions of time for $\omega_1 = 3\omega_0$. The product of these variances is also plotted. 
The dashed horizontal line is the Heisenberg limit $(\Delta Q_{\lambda})^2=1/4$ and the solid horizontal line is the coherent limit.}
 \label{varjump1}
\end{figure}

In Fig.\ref{varjump1}, we plot variances of the quadrature operators $(\Delta Q_{\lambda=0})^2$ and $(\Delta Q_{\lambda=\pi/2})^2$ as function of time for $\omega_1 = 3\omega_0$. 
Note that $(\Delta Q_{\lambda=0})^2$ is always below the coherent limit $1/2$ (solid horizontal line) and therefore is squeezed. 
Since the Heisenberg principle cannot be violated, the other quadrature is always greater than $1/2$ so that the product of both quadratures is never smaller than $1/4$ (dashed horizontal line). It is interesting that at the instant of maximum squeezing the state has a minimum uncertainties product, \textit{i.e.}, is in the Heisenberg limit.

A comment is in order here. From section \ref{IntrotoSS} it is known that, evolving a squeezed state with a TEO of the same hamiltonian, introduces dynamics in the squeezed state only through the SPh. So one could naively expect that the variance associated to a given quadrature, say for instance $(\Delta Q_{\lambda=0})^ 2$, should evolve in time in such a way that it would oscillate around the coherent limit as the spinning of the oval in Figure \ref{fig:charuto} suggests. In that case, $r$ remains constant and, since the SPh $\varphi$ is continuously increasing, this generates these oscillations. However, if we have a squeezed state relative to the hamiltonian $\hat{H}_{0} $ but evolving in time with the TEO $e^{-i{\hat H}_1 t}$, with $\hat H_1 = \frac{1}{2}{\hat p}^2 + \frac{1}{2}\omega_1^2{\hat q}^2$, as it is the case discussed in this subsection, both $r$ and $\varphi$ depend on time and the time evolution of the variances of the quadratures is more subtle, as shown in Fig.\ref{varjump1}.

\vskip 0.5cm
\centerline{\bf Interval 2}

Let us now  compute the state of the  HO at an arbitrary  instant  $t > \tau$, that is, after its frequency jumps back to its initial value $\omega_0$. 

\begin{figure}[h!]
\begin{center}
\includegraphics[width=0.5\linewidth]{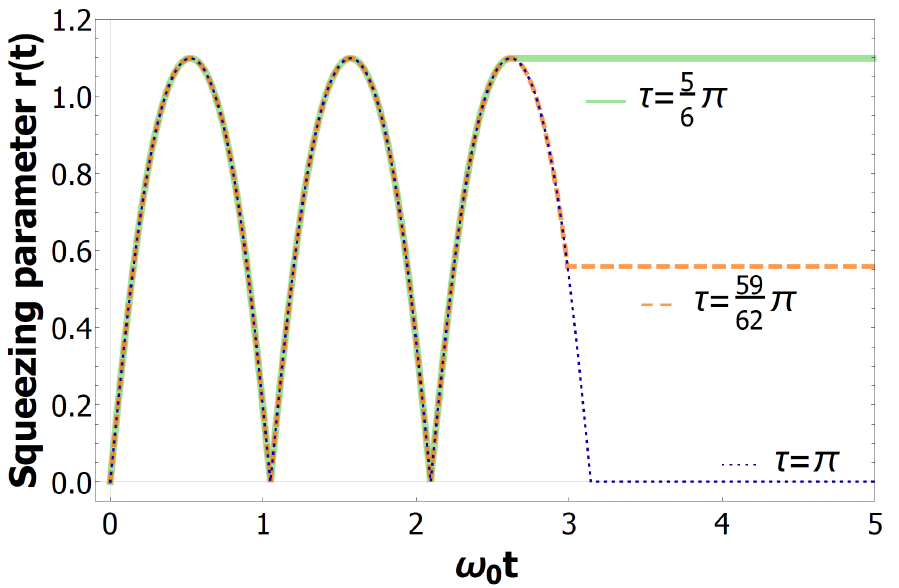}
\end{center}
\vskip -0.5cm
\caption{Squeezing parameter as a function of time for a frequency that undergoes an abrupt change at $t=0$ from $\omega_{0}=1$ (in arbitrary units) to $\omega_{1}=3\omega_0$ and a second abrupt change 
at $t=\tau$ from $\omega_1$ back to $\omega_0$, for different values of $\tau$.}
 \label{SPfixed}
\end{figure}

Using Eqs. (\ref{eq:wavefunck}), (\ref{Eq:rules}) and (\ref{eq:w1final}) it can be shown that 
\begin{eqnarray}
\left|\psi(t) \right\rangle &=& \hat U_2(t,\tau)  \vert \psi(\tau)\rangle\ =  e^{\lambda_4 \hat K_3} \vert \psi(\tau)\rangle\cr\cr
&=&
\sqrt{\left|\Lambda_{3}\right|^{1/2}}\sum_{n=0}^{\infty}\frac{\sqrt{(2n)!}}{n!}
\left[\frac{1}{2}\left|\Lambda_{+}\right|e^{i\vartheta}\right]^{n}
e^{\lambda_{4}(n+1/4)}\left|2n\right\rangle\cr\cr
&=&
\sqrt{\left|(\Lambda_{3})\right|^{1/2}}\sum_{n=0}^{\infty}\frac{\sqrt{(2n)!}}{n!}
\left[\frac{1}{2}\left|\Lambda_{+}\right|e^{i\left(\vartheta-2\omega_{0}(t-\tau)\right)}\right]^{n}\left|2n\right\rangle\, ,
\label{eq:wfinal}
%\label{eq:wf}
\end{eqnarray}
where the overall phase has been removed again and we have used Eq. (\ref{eq:lambdaspeques}). 
A few comments are in order here. First, we recall that in this interval the time evolution of the HO is described by  
the hamiltonian with the initial frequency $\omega_0$, namely, $\hat H_2 = \frac{1}{2}{\hat p}^2  +  \frac{1}{2}\omega_0 ^2{\hat q}^2$. Therefore, as we have shown 
before (see Eq. (\ref{eq:ssevol})), the SP must remain constant for $t\geq\tau$ with its value $r(\tau)$. 
In Fig.\ref{SPfixed} we plot the evolution of the SP, with its oscillatory behavior in interval 1 and its  constant value in interval 2 for
three different values of $\tau$: one of them leading to the maximum possible value for the SP ($\tau=\frac{5}{6}\pi$), 
another one with an intermediate value for the SP ($\tau=\frac{59}{62}\pi$), 
and a third one with a vanishing value for the SP ( $\tau=\pi$)

\begin{figure}[h!]
\begin{center}
\includegraphics[width=0.64\linewidth]{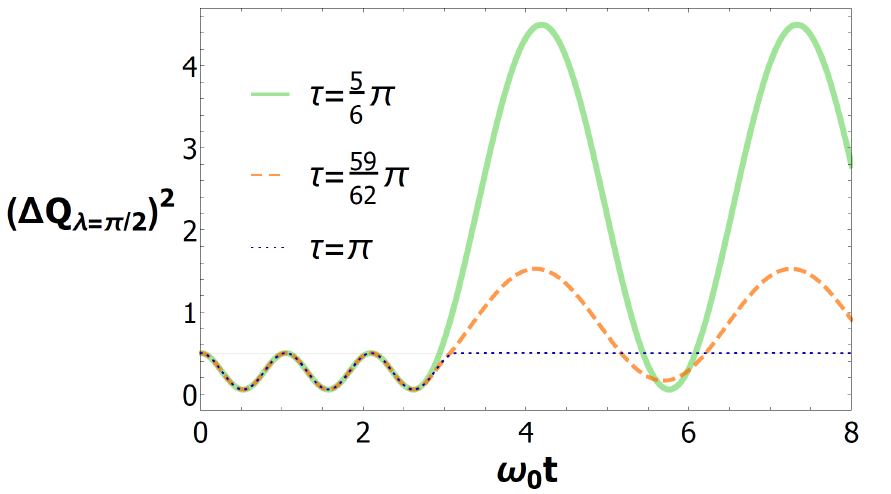}
\end{center}
\vskip -0.5cm
\caption{Plot of the variance $(\Delta Q_{\lambda=\pi/2})^2$ as a function of time  for a frequency that undergoes an abrupt change at $t=0$ from $\omega_{0}=1$ (in arbitrary units) to $\omega_{1}=3\omega_0$ and a second abrupt change 
at $t=\tau$ from $\omega_1$ back to $\omega_0$ for different values of $\tau$. The values  of $\tau$ are the same as those used in Fig.\ref{SPfixed}.}
 \label{Vardiftau}
\end{figure}

A second comment concerns the time evolution of the variances of the quadrature operators. As we have shown before,
evolving the system with the hamiltonian with the initial frequency $\omega_0$ introduces dynamics in the squeezed 
state only through the SPh (see Eq.(\ref{eq:quadratureopvariance2})). 
In fact, defining $r_\tau := r(\tau)$ and $\varphi_\tau := \varphi(\tau)$, for any instant $t>\tau$ the vector state of the HO is then given by
\begin{equation}
\left| \psi(t)\right\rangle = e^{-i\frac{1}{2}\omega_0 (t - \tau)} \sqrt{\mbox{sech}(r_\tau) }\sum_{n=0}^{\infty}\frac{\sqrt{(2n)!}}{n!}
\left[-\frac{1}{2}e^{i(\varphi_\tau - 2\omega_0(t-\tau))}\tanh (r_\tau) \right]^{n}\left|2n\right\rangle.
\label{SState-t-Tau}
\end{equation}
so that, from Eqs. (\ref{eq:quadratureopvariance2}) and (\ref{SState-t-Tau}) we see that 
\begin{equation}
(\Delta Q_\lambda)^2(t) = \frac{1}{2}\left\{ e^{2r_\tau} \sin^2\left[ \lambda -\frac{\varphi_\tau}{2} + \omega_0 (t - \tau)\right] + 
e^{-2r_\tau} \cos^2\left[ \lambda -\frac{\varphi_\tau}{2} + \omega_0 (t - \tau)\right]\right\} 
\end{equation}
An inspection in the previous formula shows that, for $t>\tau$, the variance of any quadrature operator will oscillate between a minimum value, given by $e^{-2r_\tau}/2$ 
and a maximum one, given by $e^{2r_\tau}/2$. This behaviour is shown in Fig. \ref{Vardiftau} which also shows that the behaviour of $(\Delta Q_{\lambda=\pi/2})^2$ for $0 < t \le \tau$, is in agreement with what was discussed before, in Fig. \ref{varjump1}.
Depending on the value of $\tau$, the oscillations of $(\Delta Q_{\lambda=\pi/2})^2$ can have different amplitudes. As can be noted in Fig.\ref{Vardiftau}, for $\tau=\pi$ there are no oscillations at all, since for this value of $\tau$ the SP is zero and the HO is again in its fundamental state. Note that the oscillations for $t>\tau$, no matter the chosen value for $\tau$, all have the same period 
$\frac{\pi}{\omega_{0}}$, as expected.  

We finish this section by computing the  probability amplitude for the HO to undergo a transition from its fundamental state to an arbitrary energy eigenstate $\ket{n}$ after the frequency jumps abruptly from $\omega_0$ to $\omega_1$ and from $\omega_1$ back to $\omega_0$ after a time  interval $\tau$.
In other words, we shall compute $P(n)=\left|\langle n\vert U(t,0)\vert 0\rangle\right|^{2}$, with $t>\tau$. This is known in the optics language as the photon number probability distribution and it is given by \cite{Barnett-Book-1997}
\begin{equation}
\label{eq:probsq}
P(2n) = \mbox{sech}(r)\frac{(2n)!}{(n!)^{2} 2^{2n}} (\tanh{r})^{2n} \, ,\;\;\;\; P(2n+1)=0 \, .
\end{equation}
These probabilities sum to unity, as required \cite{Barnett-Book-1997}. Using equations written in (\ref{Identifications}), 
together with Eqs. (\ref{eq:lambdaspeques}), (\ref{eq:nu}), (\ref{eq:lamb+modulus}) and (\ref{eq:lamb3modu}), in Eq. (\ref{eq:probsq}),  we obtain
\begin{eqnarray}
P(2n) &=& \left|\Lambda_{3} \right|^{1/2}\frac{(2n)!}{(n!)^{2} 2^{2n}} \left|\Lambda_{+} \right|^{2n} \\
&=&\frac{(2n)!}{(n!)^{2} 2^{2n}} \frac{\left( \frac{\left.\omega_{1}\right.^{2}-\left.\omega_{0}\right.^{2}}{2\omega_{0}\omega_{1}} \right)^{2n} \sin^{2n}(\omega_{1}\tau)}{\left[1+\left( \frac{\left.\omega_{1}\right.^{2}-\left.\omega_{0}\right.^{2}}{2\omega_{0}\omega_{1}} \right)^{2} \sin^{2}(\omega_{1}\tau)\right]^{(n+1/2)}} \, .
\end{eqnarray}
Therefore, the persistence amplitude in the fundamental state is simply given by the substitution on the above equation of $n=0$, namely
\begin{equation}
\label{eq:persamp}
Z(\omega_0,\omega_1, \tau) := \vert\langle 0\vert \psi(t)\rangle\vert^2 
=
\left[1+\left( \frac{\left.\omega_{1}\right.^{2}-\left.\omega_{0}\right.^{2}}{2\omega_{0}\omega_{1}} \right)^{2} \sin^{2}(\omega_{1}\tau)\right]^{-1/2}\, ,
\;\;\;(t > \tau)\, .
\end{equation}
Note that the previous expression does not depend on time, as expected, since after the frequency jumps back from $\omega_1$ to $\omega_0$ the hamiltonian remains constant in time. Note, also, that for $\omega_0 = \omega_1$, which means no jump, $Z(\omega_0,\omega_1;\tau) = 1$, as expected. However, even for $\omega_0 \ne \omega_1$, if $\tau = l\pi/\omega_1$, with $l = 0, 1, 2, ...$, we will also have $Z(\omega_0,\omega_1; \tau) = 1$.

\begin{figure}[h!]
\centering
\includegraphics[width=0.57\linewidth]{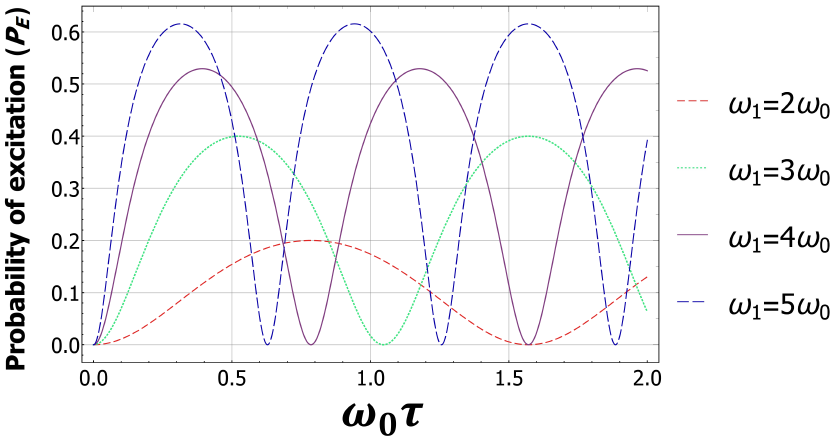}
\caption{Probability of excitation $P_{E}$ of the HO  as a function of $\tau$ (time interval in which the frequency of the HO has the value $\omega_1$) for a fixed value of $\omega_0$ and different values of $\omega_{1}$.}
\label{fig:PE2D}
\end{figure}

Once we have the persistence amplitude in the fundamental state, we can immediately write down the probability of excitation of the HO. Denoting it by 
$P_E(\omega_0,\omega_1,\tau)$, we have
\begin{equation}
P_E(\omega_0,\omega_1,\tau) = 1 - Z(\omega_0,\omega_1, \tau) \, ,
\end{equation}
which, of course, does not depend on time. In Fig.\ref{fig:PE2D} we plot $P_{E}$ as a function of $\tau$ for a fixed $\omega_0$ and different values of $\omega_1$ 
(again, without loss of generality, we have taken $\omega_{0}=1$). In this figure it is evident the  periodic behavior of $P_{E}$. 

By a direct comparison between  Fig.\ref{rVersusTau}  and  Fig.\ref{fig:PE2D} it is clear the close relation between them.
%$r(t<\tau)$ and the $P_{E}(t>\tau)$ of the HO. 
However, some care must be taken in comparing these figures, 
% \ref{rVersusTau}  and \ref{fig:PE2D}, 
%
since the former shows the time evolution of the SP $r(t)$ for $t<\tau$ while the latter shows the excitation probability  $P_{E}$ for a fixed $t$ with $t>\tau$, as a function of the 
time duration $\tau$ of the interval 1. It is evident from both figures that both, $r$ and $P_{E}$ have the same period 
%as a function of $\tau$ and $r$ as a function of $t$ have the same period, 
given by $\pi/\omega_1$, but it must be emphasized that, different from the SP, $P_{E}$ is a smooth function which has an upper bound given by 1, since it is a probability. 
% Finally, observe that, for a fixed $\tau$, different values of $\omega_1$ lead to different values of $P_{E}$. 

\begin{figure}[h!]
\begin{center}
\includegraphics[width=0.5\linewidth]{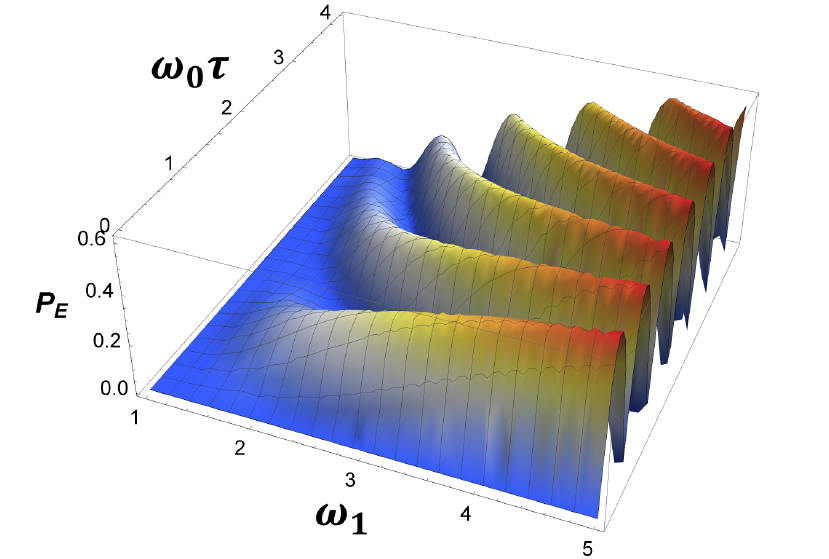}
\end{center}
\caption{Tridimensional plot of the excitation probability $P_{E}$ as a function of the time duration $\tau$  of interval 1 and the achieved frequency $\omega_1$ after the first jump.}
%Intersections of this figure with vertical planes with constant $\omega_1$ give the behaviour presented in Fig. \ref{fig:PE2D}. Analogously,  intersections of this figure with vertical planes of constant $\tau$ give the behaviour of $P_{E}$ as a function of $\omega_1$.
 \label{fig:PE3D}
\end{figure}

 We can also analyse how the excitation probability $P_{E}$, for a fixed $\tau$,  behaves as $\omega_1$ is varied.  
Although an oscillatory behaviour also appears, due to the presence of $\sin^2(\omega_1 t)$ in Eq. (\ref{eq:persamp}),  $P_{E}$ will not be a periodic function of $\omega_1$ anymore, due to the prefactor of $\sin^2(\omega_1 t)$ in this equation.  All these features can be appreciated in the \textit{3D} plot of Fig.\ref{fig:PE3D}.

\vskip 0.0cm

In Fig.\ref{fig:PE3D} the behaviour of $P_{E}$ as a function of $\omega_1$ (with fixed $\tau$) can be obtained by intersecting it with vertical planes of constant values of $\tau$. On its turn, by intersecting this figure with vertical planes of constant $\omega_1$ we reobtain the plots shown in Fig.\ref{fig:PE2D}. We finish this section by noting the resemblance of figures \ref{rVersusEtaetau} and \ref{fig:PE3D}, but their similarities must be understood on the grounds of the above comments.

%%%%%%%%%%%%%%%%%%%%%%%%%%%%%%%%%%%%%%%%%%%%%%%%%%%%%%%%%%%%%%%%%%%%%%%%%%%%%%%%%%%%%%%%%%%%%%%%%%%%%%%%%%%%%%%%%%%%%%%%%%%%%%%%%%%%%%%%%%%%%%%%%%%%%%%%%%%%%%%%%%%%%%%%%%%%%%%%%%%%%%%%%%%%%%%%%%

\section{Conclusions}\label{conclusions}

We have used algebraic methods to solve exactly a harmonic oscillator initially in the fundamental state and subjected to two sudden jumps in the frequency. It is shown that the state of the system after the first jump will evolve to a squeezed state of the initial hamiltonian. This squeezed state is characterized by having a time-dependent periodic SP. The maximum amount of squeezing \textit{i.e}., how much the variance of a quadrature is below the value for a coherent state (coherent limit), is determined by the SP. This variances of two quadratures were calculated showing periodic oscillations in time. One of the quadratures is always squeezed while the other, as the Heisenberg principle demands, is always above the coherent limit.

By a second sudden jump returning to the original frequency we have shown that the SP becomes fixed and, depending on the moment when it accounts, the SP can be maximum, zero or any possible value in between. Once the SP is fixed we shown that the variance of the squeezed quadrature begins to oscillate around the coherent limit as the typically behaviour of time-evolution of the system with the initial hamiltonian.

Finally, we have calculated the photon number probability distribution, the persistence amplitude in the fundamental state and the probability of excitation of the HO after the two jumps. We have shown how the latter is closely related to the SP by sharing the same period once is fixed the frequency of the first jump.

%%%%%%%%%%%%%%%%

%%%%%%%%%%%%%%%%

\appendix

\section{Brief derivation of Eq. \eqref{eq:BCH1}} \label{app}

\vskip 0.5cm

%Quadratic Hamiltonians are well studied in \cite{Yuen-1976}. In this appendix we show the method that enables to find the correspondent functions that equalized expressions as in eq.(\ref{eq:BCH}) for operators %closing a Lie algebra. so, we want demonstrate that, for our given set of operators $\hat{K}_{+}$,$\hat{K}_{-}$ and $\hat{K}_{3}$, satisfying the algebra
%
In this appendix we shall show that for the Lie algebras defined by the commutation relations 
\begin{equation}
\left[\hat{K}_{-},\hat{K}_{+}\right] = 2\epsilon\hat{K}_{3}\nonumber \:\:\:\:\: \mbox{and}\:\:\:\left[\hat{K}_{3},\hat{K}_{\pm}\right]=\pm\hat{K}_{\pm} \, ,
\label{eq:algebraK2}
\end{equation}
where for $\epsilon=1$ we have the $su(1,1)$ Lie algebra and for $\epsilon = -1$ we have the $su(2)$ Lie algebra, the following factorization is valid
\begin{equation}
e^{\lambda_{+}\hat{K}_{+}+\lambda_{-}\hat{K}_{-}+\lambda_{3}\hat{K}_{3}} = e^{\Lambda_{+}\hat{K}_{+}}e^{\ln(\Lambda_{3})\hat{K}_{3}}e^{\Lambda_{-}\hat{K}_{-}} \, ,
\label{eq:BCH2}
\end{equation}
where
\begin{equation}\label{Lambdas}
\Lambda_{3} = \left(\cosh(\nu)-\frac{\lambda_{3}}{2\nu}\sinh(\nu)\right)^{-2}\, ;\;\;\;\;
\Lambda_{\pm} = \frac{2\lambda_{\pm}\sinh(\nu)}{2\nu \cosh(\nu)-\lambda_{3}\sinh(\nu)} \, ,
\end{equation}
and $\nu^{2} = \frac{1}{4}\lambda_{3}^{2}-\epsilon\lambda_{+}\lambda_{-}$. 
%Note that by defining $\epsilon = 1$ we have the $su(1,1)$ Lie Algebra, while for $\epsilon = -1$ we have the $su(2)$. 
%
We will factorize these two Lie algebras simultaneously. In our demonstration,  the following BCH-like relation 
\begin{equation}
e^{\hat{A}}\hat{B}\, e^{-\hat{A}} = \hat{B}+\left[\hat{A},\hat{B}\right]+\frac{1}{2!}\left[\hat{A},\left[\hat{A},\hat{B}\right]\right]+\frac{1}{3!}\left[\hat{A},\left[\hat{A},\left[\hat{A},\hat{B}\right]\right]\right]+...\, ,
\label{BCHkey}
\end{equation}
will be used repeated times. Firstly, let us re-define the left side of Eq.(\ref{eq:BCH2}) as the special case $\theta=1$ of the operator
\begin{equation}
\hat{F}_{1}(\theta)=e^{\theta(\lambda_{+}\hat{K}_{+}+\lambda_{-}\hat{K}_{-}+\lambda_{3}\hat{K}_{3})}\, .
\label{fdt}
\end{equation}
The basic idea is to find an equivalent expression in the form
\begin{equation}
\hat{F}_{1}(\theta)=e^{\Lambda_{+}(\theta)\hat{K}_{+}}e^{\Sigma_{3}(\theta)\hat{K}_{3}}e^{\Lambda_{-}(\theta)\hat{K}_{-}}\, .
\label{fdt2}
\end{equation}
This means that we need to find write parameters $\Lambda_+$, $\Lambda_-$ and $\Sigma_3$ in terms of the small lambdas, so that the last two equations are equal. 
With this goal, we derive both expressions with respect to $\theta$ and impose the derivatives to be equal. 
%We demand the derivative of the eqs. (\ref{fdt}) and (\ref{fdt2}) respect to $\theta$ be equal. Equalizing those derivatives, join to the the fact that $\left[\hat{A},e^{\hat{A}}\right]=0$ and the unitary property of %operators of the form $e^{\hat{B}}$ i.e. $e^{\hat{B}}e^{-\hat{B}}=e^{-\hat{B}}e^{\hat{B}}=I$, 
%
Doing that and using the previous BCH formula we obtain the following coupled differential equations:
\begin{eqnarray}
\label{eqd1}
\Lambda_{+}'-\Sigma_{3}'\Lambda_{+}+\Lambda_{-}'e^{-\Sigma_{3}}\epsilon(\Lambda_{+})^{2}&=&\lambda_{+} \, ,\\
\label{eqd2}
\Sigma_{3}'-2\epsilon\Lambda_{-}'e^{-\Sigma_{3}}\Lambda_{+}&=&\Sigma_{3} \, , \\
\label{eqd3}
\Lambda_{-}'e^{-\Sigma_{3}}&=&\lambda_{-}\, \, ,
\end{eqnarray}
where the prime indicates derivative with respect to $\theta$. 
%
%Once solved this set of differential equations and then found the functions $\Lambda_{+}$, $\Sigma_{3}$ and $\Lambda_{-}$, the expression eq.(\ref{fdt2}) will be completely defined. 
%
To solve the above equations we start by decoupling them. Firstly, replacing Eq.(\ref{eqd3}) into Eqs.(\ref{eqd1}) and (\ref{eqd2}) leads to
\begin{eqnarray}
\label{coup1}
\Lambda_{+}'-\Sigma_{3}'\Lambda_{+}+\epsilon\lambda_{-}(\Lambda_{+})^{2}&=&\lambda_{+} \, , \\
\label{coup2}
\Sigma_{3}'-2\epsilon\lambda_{-}\Lambda_{+}&=&\lambda_{3} \, .
\end{eqnarray}
Then, we obtain a differential equation for the function $\Lambda_{+}$ by isolating $\Sigma_{3}'$ from Eq.(\ref{coup2}) and replacing it into Eq.(\ref{coup1}):
\begin{equation}
\boxed{
\frac{d}{d\theta}\Lambda_{+}=\lambda_{+}+\lambda_{3}\Lambda_{+}+\epsilon\lambda_{-}(\Lambda_{+})^{2}} \, .
\label{Riccati}
\end{equation}
Eq.(\ref{Riccati}) is a first order, quadratic and non-homogeneous  ordinary differential equation known as the \textsl{Riccati equation}. It has unique solution and can be transformed into an ordinary, homogeneous and second order differential equation with the aid of  the transformation 
\begin{equation}
\Lambda_{+}=-\frac{1}{\epsilon\lambda_{-}}\frac{1}{u}\frac{du}{d\theta} \, ,
\label{lambda+final}
\end{equation}
leading to
\begin{equation}
\boxed{
\frac{d^{2}u}{\left.d\theta\right.^{2}}+\Gamma\frac{du}{d\theta}+\omega_{0}^{2}u=0} \, ,
\label{gener}
\end{equation}
where we defined $\omega_{0}^2 = \epsilon\lambda_{-}\lambda_{+} \;\;\;\mbox{and}\:\:\:\:\: \Gamma = -\lambda_{3} \,$ in order to identify it as the classical equation of a damped harmonic oscillator with natural frequency $\omega_{0}$ and damped coefficient $\Gamma$. Note that by using Eqs.(\ref{coup2}) and (\ref{lambda+final}) we can calculate
\begin{equation}
\Sigma_{3}(\theta) = \lambda_{3}\theta-2\epsilon\lambda_{-}(\frac{1}{\epsilon\lambda_{-}})\int{\frac{du}{u}}+C_{1}\, =\;  \lambda_{3}\theta-2\ln{u(\theta)}+C_{1} 
\label{lambda3final}
\end{equation}
and replacing Eq.(\ref{lambda3final}) into Eq.(\ref{eqd3}) we obtain
\begin{equation}
\Lambda_{-}(\theta) = \lambda_{-}\int{e^{\Sigma_{3}(\theta)}d\theta}+C_{2} \, ,
\label{lambda-final}
\end{equation}
where constants $C_{1}$ and $C_{2}$ are determined from the initial conditions, namely, $\Sigma_{3}(\theta=0) = 0$ and $ \Lambda_{-}(\theta=0) = 0$. Therefore, once we find  $u(\theta)$ we can determine the functions $\Lambda_+$, $\Lambda_-$ and $\Sigma_3$. Coming back to Eq. (\ref{gener}), it general solution is given by
\begin{equation}
u(\theta)=e^{-\frac{\Gamma}{2} \theta}\left(Ae^{\nu\theta}+Be^{-\nu\theta}\right) \, ,
\label{genersol}
\end{equation}
where
\begin{equation}
\nu=\sqrt{\frac{1}{4}\Gamma^{2}-\omega_{0}^{2}}=\sqrt{\frac{1}{4}\lambda_{3}^{2}-\epsilon\lambda_{-}\lambda_{+}} \, 
\label{eq:nu1}
\end{equation}
and constants $A$ and $B$ are  determined from the initial conditions. Using the above results in Eq.(\ref{lambda+final}) we find
\begin{equation}
\Lambda_{+}(\theta) = -\frac{1}{\epsilon\lambda_{-}}\frac{1}{u}\left[(\nu-\frac{\Gamma}{2})u-2\nu B e^{-(\frac{\Gamma}{2}+\nu)\theta}\right]=\frac{(\frac{\Gamma}{2}-\nu)}{\epsilon\lambda_{-}} + \frac{2\nu B e^{-\nu\theta}}{\epsilon\lambda_{-}\left[Ae^{\nu\theta}+Be^{-\nu\theta}\right]},
\label{lambdafound}
\end{equation}
and from the initial condition $\Lambda_{+}(\theta=0)=0$ we get $A=\frac{(\nu+\Gamma/2)}{(\nu-\Gamma/2)}B$. Therefore Eq.(\ref{lambdafound}) becomes
\begin{eqnarray}
\Lambda_{+}(\theta) 
&=&
\frac{1}{\epsilon\lambda_{-}}\left[\left(\frac{\Gamma}{2}-\nu\right)+\frac{2\nu\left(\nu-\frac{\Gamma}{2}\right) e^{-\nu\theta}}{\left[\left(\nu+\frac{\Gamma}{2}\right)e^{\nu\theta}+\left(\nu-\frac{\Gamma}{2}\right)e^{-\nu\theta}\right]}\right]\nonumber\\
%\Rightarrow \:\:\: \Lambda_{+}
&=&
\frac{1}{\epsilon\lambda_{-}}\left[\frac{2\left(\frac{\Gamma^{2}}{4}-\nu^{2}\right)\sinh(\nu\theta)}{2\nu \cosh(\nu\theta)+\Gamma \sinh(\nu\theta)}\right].\nonumber
\label{lambdafofin}
\end{eqnarray}
Now, using definition Eq.(\ref{eq:nu1}) and expressions $\omega_0^2 = \epsilon\lambda_-\lambda_+$ and $\Gamma = -\lambda_3$, we obtain
\begin{equation}\boxed{
\Lambda_{+}(\theta)=\frac{2\lambda_{+}\sinh(\nu\theta)}{2\nu \cosh(\nu\theta)-\lambda_{3} \sinh(\nu\theta)}},\nonumber
\label{lambdamais}
\end{equation}
which leads to the desired expression written in Eq.(\ref{Lambdas})  if we take $\theta=1$.

In order to obtain $\Sigma_{3}$ we first replace the solution for $u(\theta)$,  Eq.(\ref{genersol}), into  Eq.(\ref{lambda3final}) to get
\begin{eqnarray}
\Sigma_{3}(\theta) 
&=&
\lambda_{3}\theta-2\ln{\left\{e^{-\frac{\Gamma}{2} \theta}\left(\frac{(\nu+\Gamma/2)}{(\nu-\Gamma/2)}e^{\nu\theta}+e^{-\nu\theta}\right)B\right\}}+C_{1} \nonumber\\
&=&(\lambda_{3}+\Gamma)\theta-2\ln{\left(\frac{(\nu+\Gamma/2)}{(\nu-\Gamma/2)}e^{\nu\theta}+e^{-\nu\theta}\right)} + D\, ,
\label{lambda3sol}
\end{eqnarray}
where all constants have been absorbed in $D$. Using the initial condition $\Sigma_3(0) = 0$ and that $\Gamma = - \lambda_3$ it can be shown that $ D = 2\ln{(\frac{2\nu}{\nu-\Gamma/2})}$. Replacing this result into Eq.(\ref{lambda3sol}) we get
\begin{equation}\boxed{
\Sigma_{3}(\theta) = \ln{\left\{\left(\cosh(\nu\theta)-\frac{\lambda_{3}}{2\nu} \sinh(\nu\theta)\right)^{-2}\right\}}}\, ,
\label{Lambdatres}
\end{equation}
which after taking $\theta = 1$ leads to the desired result of equation in (\ref{Lambdas}) since in Eq.(\ref{eq:BCH2}) $\Lambda_{3}$ is defined as the argument of the logarithm.

Finally, in order  to find $\Lambda_{-}(\theta)$ we  replace Eq.(\ref{Lambdatres}) into Eq.(\ref{lambda-final}) obtaining
\begin{eqnarray}
\Lambda_{-}(\theta)
&=&
\lambda_{-}\int{\frac{\mbox{sech}^{2}(\nu\theta)}{\left(1-\frac{\lambda_{3}}{2\nu} \mbox{tanh}(\nu\theta)\right)^{2}}d\theta}+C_{2}\nonumber\\
&=&
\frac{2\lambda_{-}}{\lambda_{3}}\left(\frac{2\nu \cosh(\nu\theta)}{2\nu \cosh(\nu\theta)-\lambda_{3} \sinh(\nu\theta)}\right)+C_{2}.
\label{Lambdamenos}
\end{eqnarray}
Using the initial condition for $\Lambda_{-}$ it can be shown that $C_{2} = -\frac{2\lambda_{-}}{\lambda_{3}}$. Substituting this result in Eq.(\ref{Lambdamenos}) we finally obtain
\begin{equation}\boxed{
\Lambda_{-}(\theta) = \frac{2\lambda_{-}\sinh(\nu\theta)}{2\nu \cosh(\nu\theta)-\lambda_{3} \sinh(\nu\theta)}},
\label{Lfin}
\end{equation}
which leads to the desired result after taking $\theta = 1$.

\begin{acknowledgments}
The authors acknowledge Reinaldo F. de Melo e Souza, M. V. Cougo-Pinto and P.A. Maia Neto for enlightening discussions. The authors thank the Brazilian agencies for scientific and technological research CAPES, CNPq and FAPERJ for partial  financial support.
\end{acknowledgments}

%%%%%%%%%%%%%%%%%%%%%%%%%%%%%%%%%%%%%%%%%%%%%%%%%%%%%%%%%%%%%%%%%%%%%%%%%%%%%%%%%%%%%%%%%%%%%%%%%%%%%%%%%%%%%%%%%%%%%%%%%%%%%%%%%%%%%%%%%%%%%%%%%%%%%%%%%%%%%%%%%%%%%%%%%%%%%%%%%%%%%%%%%%%%%%%%%%

\end{document}